\documentclass[%
 reprint,
 amsmath,amssymb,
 aps,
]{revtex4-1}

\usepackage{graphicx}
\usepackage{dcolumn}
\usepackage{bm}

\begin{document}


\title{Entropic $f(R)$ Gravity}

\author{Ali Teimouri}
\email{a.teimouri@lancaster.ac.uk}
\email{ilia.teimouri@gmail.com}
\affiliation{Consortium for Fundamental Physics, 
\\Lancaster University, Lancaster, LA$1$ $4$YB, United Kingdom.}


\begin{abstract}
In this short paper we follow the entropic gravity approach and demonstrate
how $f(R)$ theories of gravity can be emergent. This is done by introducing
an effective gravitational constant which is naturally arising from the $f(R)$'s
equations of motion. 
\end{abstract}

\maketitle


\section{\label{sec:level1}Introduction}
Gravitation is a universal force which interacts with all particles that carry an energy and thus there is a link between gravity and thermodynamics. The understanding of this relation is matured in the past decades by studying the black holes' thermodynamics. For instance, Jacobson  \cite{Jacobson:1995ab} derived the Einstein's equations from the thermodynamics of the near horizon. Another important advancement in understanding the relation between thermodynamics and gravity was achieved by studying the black holes' entropy, where Hawking and Bekenstein \cite{refref} have shown that this entropy  is proportional to the area of the event horizon. More recently, and inspired by the area law, the holographic principle \cite{Susskind:1994vu} was realised in the context of $AdS/CFT$ correspondence \cite{Maldacena:1997re}.

Inspired by these developments, there is a new conjecture that sees gravity not as a fundamental force but as an emerging phenomenon. Examples of this approach can be found in \cite{Verlinde:2010hp,Padmanabhan:2009kr}. Particularly, in \cite{Verlinde:2010hp}, the gravity is  thought to be an entropic force, where the gravitational formulation can be derived by obtaining a relation between temperature and acceleration, and then using the  holographic principle and an equipartition rule relating the energy to the temperature and the number of degrees of freedom.

Gravity, as we know it so far from the   Einstein's theory of general relativity, is fairly successful in predicting the natural phenomena. One of the most recent phenomenon was the prediction of the gravitational waves.  However, the limits of general relativity brought  the alternative theories of gravity to existence. Many of these alternative theories are essentially wide range of modifications to the original general relativity. From these modifications, one can mention the $f(R)$ gravity \cite{Sotiriou:2008rp}, which generalises the Einstein's theory of general relativity and was used most famously by Starobinsky \cite{Starobinsky:1980te}, to describe the cosmic inflation. There are also many other types of modified theories of gravity \cite{Clifton:2011jh}, each constructed to explain different phenomena when the general relativity is not able to provide an appropriate description. For instance, one can recall Lovelock, Gauss-Bonnet, higher derivative gravity \cite{Biswas:2011ar} and so on. 

Looking at the gravity as an emergent phenomenon, and deriving the Einstein's theory of general relativity from the basic thermodynamics, raises a question about the modified theories of gravity. Is it possible to derive the modified theories of gravity from the basic laws of thermodynamics? How can the modified theories of gravity be emerged from the basic laws and how do these theories reflect themselves in those basic laws? In this paper we wish to show how the  $f(R)$ gravity can be obtained from the basic principles. We are going to do so by implementing Verlinde's approach in \cite{Verlinde:2010hp}. 

We start by giving a brief review of the entropic gravity and we then move onto the emergence of the $f(R)$ gravity.   
           
\section{\label{sec:level1}Entropic Gravity}
The existence of the area law in general relativity implies that the space-time is nothing but a perfect storage for information and that this information can be read off from the boundary which is defined by the area law. The information stored on the area is maximal and thus finite. This is to satisfy the Bekenstein's entropy bound \cite{Bekenstein:1980jp}. In similar analogy, it can be assumed that the information is stored on the so called  \textit{screens}. Screens separate points and therefore one can locate the stored particles in discrete bits on the screen.  

Verlinde, \cite{Verlinde:2010hp}, argued that it is possible to use the second
law of thermodynamics and derive the Einstein's theory of general relativity
from the first principle. This had been done by considering a holographic
screen on closed surface of constant redshift. The assumption is that there is
a associated mass configuration to the screen with total mass $M$. The bit density
on the screen is then simply:
\begin{equation}\label{bitden}
dN=\frac{dA}{G\hbar},
\end{equation}
where $N$ is a number of bits, $A=4\pi r^{2}$ is the surface area, $G$ is
the gravitational constant and $\hbar$ is the Planck's constant. Given that the
total energy of the system is denoted by $E$, the temperature can be determined
by the equipartition principle as, \begin{equation}\label{energy1}
E=\frac{1}{2}Nk_B T,
\end{equation}
where $k_B $ is the Boltzman's constant. We can use the mass-energy equivalence
and drop out the constants appropriately, and thus determine the mass
that each bit carries by simply integrating the mass, which is: 
\begin{equation}
M=\frac{1}{2}\int_{\sigma}T dN=\frac{1}{4\pi}\int_{\sigma}\nabla\varphi dA.
\end{equation} 
The above equation is known as Gauss' law for gravity and can be re-expressed in terms of the Komar mass which is sufficient to derive the Einstein's equations.

\section{$f(R)$ Gravity Emergence}
$f(R)$ gravity is the generalisation of the Einstein's theory of general relativity. 
The action can be written in the form of \cite{Sotiriou:2008rp}:
\begin{equation}
S=\frac{1}{16\pi G}\int dx^{4}\sqrt{-g}f(R)+S_{M}(g_{\mu\nu},\psi),
\end{equation}
where $f(R)$ is the function of scalar curvature, $R$,  $S_{M}$ denotes the matter term with $\psi$ being the matter fields.  The
variation of the action with respect to the metric gives: 
\begin{equation}\label{eom}
f'(R)R_{\mu\nu}-\frac{1}{2}f(R)g_{\mu\nu}-[\nabla_{\mu}\nabla_{\nu}-g_{\mu\nu}\Box]f'(R)=8\pi
G T_{\mu\nu}.
\end{equation}
We can re-write above as: 
\begin{equation}
G_{\mu\nu}=\frac{8\pi G}{f'(R)}\Big(T_{\mu\nu}+T^{(eff)}_{\mu\nu}\Big),
\end{equation}
where we  introduced the effective gravitational coupling strength as,
 \begin{equation}
G_{eff}\equiv\frac{G}{f'(R)}. 
\end{equation}
Introducing the effective gravitational constant $G_{eff}$ is equivalent
to the requirement that the graviton is not a ghost.  Moreover, $T^{(eff)}_{\mu\nu}$ is called the effective stress-energy tensor and can be easily read off from Eq. (\ref{eom}). 
\subsection{Newtonian Limit}
In this setup, let us consider the weak-field approximation. The metric describing
 the gravitational field of a static distribution of matter
is given by, 
\begin{equation}\label{met}
ds^2=-(1+2\varphi)dt^{2}+(1-2\varphi)d\bar x^{2},
\end{equation}
where $d\bar x^{2}=dx^{2}+dy^{2}+dz^{2}$ and $\varphi(r)$ is the Newtonian potential and it is the function of distance $r$. In asymptotically flat space-time,
the weak-field expression given above, can be used to approximate the metric
in the asymptotic domain. At far distance from the static gravitating object
we have \cite{frolov}, 
\begin{equation}
\varphi(r)=-\frac{G_{eff}M}{r}. 
\end{equation}
Thus, the free-fall acceleration can be obtained by taking the gradient of the Newtonian potential: 
\begin{equation}
a^{i}=-\nabla\varphi(r)=-\frac{G_{eff}M}{r^{2}}n^{i},
\end{equation}
where $n^{i}$ is a unit vector of the external normal to a 2D sphere $\sigma$
of radius $r$. The mass of the object  can then be found via: 
\begin{equation}\label{newtonianmass}
M=\frac{1}{4\pi G_{eff}}\int_{\sigma}a^{i}n_{i}d^{2}\sigma.
\end{equation}
Again, this is the familiar Gauss law for gravity.
\\
\subsection{Komar mass}
It is possible to write Eq. (\ref{newtonianmass}) in terms of the Komar mass by identifying the Killing vector associated to the metric in Eq. (\ref{met}),
\cite{komar}, \begin{equation}
M=\frac{1}{4\pi G_{eff}}\int_{\sigma}\nabla^{\nu}\xi^{\mu}d\sigma_{\mu\nu},
\quad\text{with:}\quad d\sigma_{\mu\nu}=n_{[\mu}u_{\nu]}d^{2}\sigma.
\end{equation}
In the above definition of mass, $\xi^{\mu}$ is the Killing vector field of a
static space-time. Moreover, $n_{\alpha}$ and $u_{\alpha}$ are the time-like and space-like normals to $\sigma$. By using the cyclic identity for Riemann tensor, and the
fact that all Killing vectors must satisfy
\begin{equation}
\Box\xi^{\alpha}=-R^{\alpha}_{\ \beta}\xi^{\beta},
\end{equation}
and also by using the Stokes' theorem, one has: 
\begin{equation}
M=\frac{1}{4\pi G_{eff}}\int_{\partial\Sigma}\nabla^{\nu}\xi^{\mu}d\sigma_{\mu\nu}=-\frac{1}{4\pi
G_{eff}}\int_{\Sigma}R^{\mu}_{\ \nu}\xi^{\nu}d\sigma_{\mu}.
\end{equation}
Here, $\Sigma$ is a 3-dimensional volume bounded by holographic boundary
$\partial\Sigma$. We shall note that $d\sigma_{\mu}$ is proportional to the normal to $\Sigma$. It can be clearly seen that upon expanding the effective
gravitational constant and taking the example of Einstein-Hilbert action, which
is $f(R)=R$, one recovers the results for general relativity. 
\subsection{Emergence of gravity}
As we saw from Eq. (\ref{bitden}),  in order to derive gravity from entropy
one has to start with the bit density on the holographic screen. In the example
of the $f(R)$ gravity this shall be modified to, 
\begin{equation}
dN=\frac{dA}{G_{eff}\hbar}.
\end{equation}
Again, the boundary can be thought as a surface where the information is stored. Upon satisfying the holographic principle, the maximal storage space (\textit{i.e.} the total number of bits) is proportional to the area, $A$. It is now possible to  repeat the same procedure to essentially  find the total energy related to the number of bits and the temperature as in Eq. (\ref{energy1}) and then find the associated mass in terms of Komar integral and satisfy the equations of motion.
\section{Summary}
In this paper, we have shown that one can obtain the $f(R)$ theories  of gravity by using the entropic analogy of gravity. This requires introducing an effective gravitational constant in the Newtonian potential. This effective gravitational constant comes immediately from the equations of motion. This explains how the modification of the original theory of general relativity affected the Newtonian potential. 

It is clear that the same approach can be employed to derive other theories of modified gravity. This is due to the fact that it is possible to generalise the Komar integral for higher order terms. However, the quest to recognise the most appropriate theory of gravity remains open to study.
 
\section*{Acknowledgement}
The author would like to thank Spyridon Talaganis for fruitful discussions
and comments. 




\begin{thebibliography}{99}
\bibitem{Jacobson:1995ab} 
  T.~Jacobson,
  Phys.\ Rev.\ Lett.\  {\bf 75}, 1260 (1995)
  doi:10.1103/PhysRevLett.75.1260
  [gr-qc/9504004].

\bibitem{refref}
J. M. Bardeen, B. Carter and S. W. Hawking, "The Four laws of black hole mechanics," Commun. Math. Phys. 31, 161 (1973). J. D. Bekenstein, "Extraction of energy and charge from a black hole," Phys. Rev. D 7, 949 (1973). J. D. Bekenstein, "Black holes and entropy," Phys. Rev. D 7, 2333 (1973). S. W. Hawking, "Particle Creation By Black Holes," Commun. Math. Phys. 43, 199 (1975) [Erratum-ibid. 46, 206 (1976)].

\bibitem{Susskind:1994vu} 
  L.~Susskind,
  J.\ Math.\ Phys.\  {\bf 36}, 6377 (1995)
  doi:10.1063/1.531249
  [hep-th/9409089].  G.~'t Hooft,
  Salamfest 1993:0284-296
  [gr-qc/9310026].

\bibitem{Maldacena:1997re} 
  J.~M.~Maldacena,
  Int.\ J.\ Theor.\ Phys.\  {\bf 38}, 1113 (1999)
  [Adv.\ Theor.\ Math.\ Phys.\  {\bf 2}, 231 (1998)]
  doi:10.1023/A:1026654312961
  [hep-th/9711200].

\bibitem{Verlinde:2010hp} 
  E.~P.~Verlinde,
  JHEP {\bf 1104}, 029 (2011)
  doi:10.1007/JHEP04(2011)029
  [arXiv:1001.0785 [hep-th]].
  
\bibitem{Padmanabhan:2009kr} 
  T.~Padmanabhan,
  Mod.\ Phys.\ Lett.\ A {\bf 25}, 1129 (2010)
  doi:10.1142/S021773231003313X
  [arXiv:0912.3165 [gr-qc]].

\bibitem{Sotiriou:2008rp} 
  T.~P.~Sotiriou and V.~Faraoni,
  Rev.\ Mod.\ Phys.\  {\bf 82}, 451 (2010)
  doi:10.1103/RevModPhys.82.451
  [arXiv:0805.1726 [gr-qc]].
  
\bibitem{frolov}
V.P. Frolov,  A. Zelnikov, Introduction to Black Hole Physics (Oxford University
Press, New York, 2011), p. 138.

\bibitem{komar}
A. Komar, Covariant Conservation Laws in General Relativity, Phys. Rev. 113,
934  (1959)

\bibitem{Starobinsky:1980te} 
  A.~A.~Starobinsky,
  Phys.\ Lett.\  {\bf 91B}, 99 (1980).
  doi:10.1016/0370-2693(80)90670-X
  
\bibitem{Clifton:2011jh} 
  T.~Clifton, P.~G.~Ferreira, A.~Padilla and C.~Skordis,
  Phys.\ Rept.\  {\bf 513}, 1 (2012)
  doi:10.1016/j.physrep.2012.01.001
  [arXiv:1106.2476 [astro-ph.CO]].
  
\bibitem{Biswas:2011ar} 
  T.~Biswas, E.~Gerwick, T.~Koivisto and A.~Mazumdar,
  Phys.\ Rev.\ Lett.\  {\bf 108}, 031101 (2012)
  doi:10.1103/PhysRevLett.108.031101
  [arXiv:1110.5249 [gr-qc]].
  
\bibitem{Bekenstein:1980jp} 
  J.~D.~Bekenstein,
  Phys.\ Rev.\ D {\bf 23}, 287 (1981).
  doi:10.1103/PhysRevD.23.287
\end{thebibliography}
\end{document}